\documentclass[showpacs,prb,twocolumn]{revtex4}

\usepackage[dvips]{graphicx}
\usepackage{amsmath}

\date{\today}

\begin{document}

\pacs{72.25.Rb, 72.70.+m, 73.23.Hk, 73.63.Kv}

\title{Single Spin Transport Spectroscopy -  Current Blockade and Spin Decay}

\author{Gerold Kie{\ss}lich}
\email{kiesslich@itp.physik.tu-berin.de}
\author{Gernot Schaller}
\author{Clive Emary}
\author{Tobias Brandes}

\keywords{Heisenberg exchange coupling, single-electron transistor, sequential tunneling}

\affiliation{Institut f{\"u}r Theoretische Physik, Hardenbergstr. 36, Technische Universit{\"a}t
  Berlin, D-10623 Berlin, Germany}

\begin{abstract}
We present a theory of a single-electron transistor exchange-coupled
to a localized spin. We show how to gain detailed quantitative
knowledge about the attached spin such as spin size, exchange coupling
strength, Land\'e g-factor, and spin decay time $T_1$ by utilizing a robust 
blockade phenomenon 
of DC magnetotransport with accompanying noise enhancement. 
Our studies are of particular relevance to
spin-resolved scanning single-electron transistor microscopy,
electronic transport through nanomagnets, and
the effect of hyperfine interaction on
transport electrons by surrounding nuclear spins.
\end{abstract}

\maketitle


The selective detection and manipulation of single spins are of fundamental importance for the realization of qubits as basic building-blocks
in quantum computers \cite{LOS98}. In a solid-state environment,
spins experience relaxation and decoherence which limit the functionality of quantum information processing.
Therefore it is necessary to shed light  onto the time scales of those
effects. In addition, properties such as the coupling between spins as
well as to an external magnetic field are important. Recently, the electrical time-resolved read-out of an individual electron spin enabled the measurement 
of the relaxation time \cite{ELZ04} $T_1$. Furthermore, 
Wabnig {\em et al.} have proposed obtaining the $T_1$ and $T_2$ times 
by high-frequency (GHz) noise measurements \cite{WAB09}.
Such experiments would be very challenging and we suggest here 
an alternative route to access at least $T_1$ which avoids time-resolved or high-frequency setups.

In this letter we explore the nonlinear current and low-frequency noise of a single-electron transistor (SET) where the spin of the quantum dot (QD) electron is 
coupled via the exchange interaction to a second localized spin. We find
that tunneling spectroscopy in a constant magnetic field reveals a
robust current blockade region due to population 
trapping  that allows for the determination of $T_1$ in a simple
manner. The proposed setup could be used as a spin-resolved version of
the scanning single-electron transistor (SSET) \cite{YOO97}, as alternative to single-spin detection by magnetic resonance force microscopy (MRFM) \cite{RUG04}.

There are two further situations for which our work is of relevance:
Electronic transport through nanomagnets (e.g. QDs doped with Mn-ions
\cite{ROS07}) 
with an attached spin of 5/2 and an anisotropic
exchange coupling; and as a starting point for studying the
effect of hyperfine interaction on
transport electrons by surrounding nuclear spins in QDs (e.g. as shown in Refs.~\cite{ONO04,INA07a} for double-QDs).


\begin{figure}[b]
\includegraphics[width=.48\textwidth]{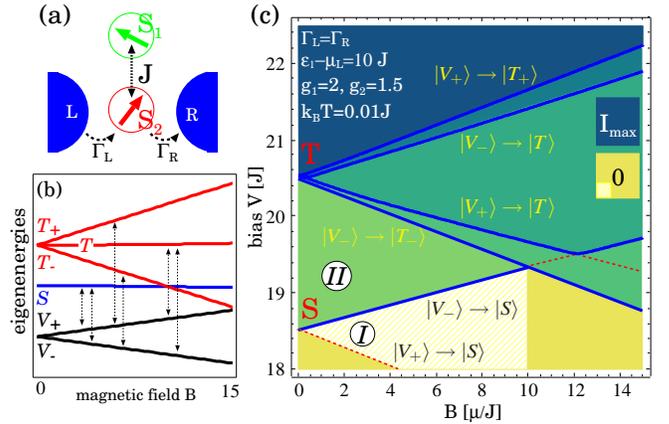}
\caption{(color online). (a) Sketch of the single-electron transistor (SET) with lead
couplings $\Gamma_{L,R}$ and electronic level with spin $\vec{S}_2$
and attached spin $\vec{S}_1$ via exchange coupling $J$. (b)
Eigenenergies of $H_{\textrm{QD}}$ (singlet $S$, triplets $T, T_\pm$, unoccupied
SET-level $V_\pm$) {\em vs.} magnetic field $B$ for $g_{1,2}>0$ (see
Tab.~\ref{tab1}). Arrows indicate allowed transitions due to single-electron
tunneling. (c) Steady-state SET current {\em vs.} symmetric bias voltage $V$
and magnetic field $B$; Solid lines: current steps. Dotted lines:
blocked transitions; Shaded region $I$: current blockade, bunched
electron transfer (see also Fig.~\ref{fig2}a). Region $II$: see text.}
\label{fig1}
\end{figure}

Our model consists of a SET with constant lead couplings
$\Gamma_{L,R}$ and a QD electron spin $\vec{S}_2$
which is isotropically exchange-coupled to an additional spin
$\vec{S}_1$ with strength $J$ as sketched in Fig.~\ref{fig1}a.
The corresponding Hamiltonian (for the exchange term see also Ref.~\cite{GOL04}), essentially an Anderson Model with exchange-coupled
spin $\vec{S}_1$, reads
\mbox{$H=H_\textrm{QD}+H_\textrm{leads}+H_\textrm{T}$}, \mbox{$H_\textrm{leads}=\sum_{k\sigma\alpha
=L,R}\varepsilon_{k\alpha}
c^\dagger_{k\alpha\sigma}c_{k\alpha\sigma}$},

\begin{eqnarray}
\label{eq:hamilt}
H_\textrm{QD}&=&\sum_{i=1,2}\big(\varepsilon_i n_i+ \mu g_i
\vec{B}\cdot\vec{S_i}\big) +J\vec{S_1}\cdot\vec{S_2},\\
H_\textrm{T}&=&\sum_{k\alpha\sigma_2}\big(t_{k\alpha\sigma_2}c^\dagger_{k\alpha\sigma_2}d_{\sigma_2}
+\textrm{h.c.}\big),\nonumber
\end{eqnarray}
with 
$\varepsilon_i$ the single-particle energy of the QD ($i=2$)
and the attached spin level ($i=1$), $n_i=\sum_{\sigma_i}
d_{\sigma_i}^\dagger d_{\sigma_i}$. Throughout this work the SET operates
in the Coulomb blockade regime, i.e., only a single electron can enter
the SET. This restricts the dimension of the accessible Hilbert space.
The attached state with spin $\vec{S_1}$ is always occupied
with one electron.  
The Zeeman-energies
$\mu g_i
\vec{B}\cdot\vec{S_i}$ contain the Land\'e factors
$g_i$ and magnetic field $\vec{B}$. $J$ denotes the
exchange coupling strength. The operators $d_i^\dagger/d_i$ ($i=1,2$) are the
creation/annihilation operators in the attached spin/SET, and 
$c_{k\alpha\sigma}^\dagger/c_{k\alpha\sigma}$ are the
creation/annihilation operators in lead $\alpha=L,R$ for momentum $k$
and spin $\sigma$.

For the sake of simplicity and clarity we consider a spin $S_1=1/2$ throughout this work.
The corresponding eigenenergies and -states of $H_\textrm{QD}$ are presented in
Tab.~\ref{tab1}, with the magnetic field-dependent spectrum shown in
Fig.~\ref{fig1}b.  
Note that the coefficients $c_i$ ($i=1\dots 4$) depend on the
magnetic field. 
The Coulomb interaction between the QD electron and the 
side-electron simply produces an overall-shift of the spectrum and is
therefore not considered further.

\begin{table}[t]
\begin{tabular}{|l|l|}
\hline
$| V_+\rangle\equiv| \uparrow 0\rangle$ & $\varepsilon_1 + \varepsilon_z^{(1)}/2$\\
$| V_-\rangle\equiv| \downarrow 0\rangle$ & $\varepsilon_1 - \varepsilon_z^{(1)}/2$\\
\hline
$| T_+\rangle\equiv| \uparrow\uparrow\rangle$ & $\varepsilon + \varepsilon_z/2 + 1/4$\\
$| T_-\rangle\equiv| \downarrow\downarrow\rangle$ & $\varepsilon - \varepsilon_z/2 + 1/4$\\
$| T\rangle\equiv\frac{1}{c_1}\bigg\{c_3|\uparrow\downarrow\rangle +|\downarrow\uparrow\rangle \bigg\}$ & $\varepsilon - 1/4 + c/2$\\
$| S\rangle\equiv\frac{1}{c_2}\bigg\{c_4|\uparrow\downarrow\rangle +|\downarrow\uparrow\rangle \bigg\}$ & $\varepsilon - 1/4 - c/2$\\
\hline
\end{tabular}
\caption{Eigenenergies and -states of $H_\textrm{QD}$ (\ref{eq:hamilt}) for
$\vec{B}=B\vec{e}_z$, $\varepsilon_z^{(i)}\equiv \mu g_i B/J$, $\varepsilon\equiv\varepsilon_1+\varepsilon_2$, $\varepsilon_z\equiv\varepsilon_z^{(1)}+\varepsilon_z^{(2)}$, $\Delta\varepsilon_z\equiv\varepsilon_z^{(1)}-\varepsilon_z^{(2)}$, 
$c_{1/2}\equiv \sqrt{1+c_{3/4}^2}$, $c_{3/4}\equiv \Delta\varepsilon_z\pm c$,
$c\equiv\sqrt{4\Delta\varepsilon_z^2+1}$. (All energies are scaled by
the exchange coupling strength $J$ throughout the paper. Realistic values for $J$ can be estimated, e.g., by the dipole-dipole interaction between two electron spins\cite{WAB09a} 
$J\approx\big(r_0/r\big)^3\,0.1\mu$eV with distance $r$ and $r_0=1$nm. )} 
\label{tab1}
\end{table}

We treat the coupling to the contacts in second-order perturbation
theory (sequential tunneling approximation).
The standard Born-Markov-Secular approximation \cite{BRE02,SCH08} for
$B>0$ (non-degenerate spectrum) leads to
rate equations   in the
energy-eigenbasis. They can be cast into the compact form $\dot\rho
=\mathcal{L}\rho$ with \mbox{$\rho =\big(\rho_{(\uparrow 0)},\rho_{(\downarrow 0)},
\rho_{T_+},\rho_{T_-},\rho_T,\rho_S\big)^T$} and
$\mathcal{L}\equiv\mathcal{L}_L+\mathcal{L}_R$.
The Liouvillians (6$\times$6-matrices) $\mathcal{L}_{L/R}$ contain the
couplings $\Gamma_{L/R}^\sigma =2\pi\sum_k|t_{kL/R\sigma}|^2\delta (\varepsilon -\varepsilon_{kL/R\sigma})$, 
the Fermi functions $f_{L/R}(\epsilon_i )$ for the
occupation of the $L/R$-contacts with the excitation energy
$\epsilon_i$ and respective bias voltage $\pm V/2$, and the
Clebsch-Gordan coefficients $c_3 /c_1$, $1/c_1$, $c_4 /c_2$, $1/c_2$ (see Tab.~\ref{tab1}).
Counting fields $e^{\pm i\chi}$ are included in the right lead
Liouvillian $\mathcal{L}_R =\mathcal{L}_R(\chi )$ \cite{BAG03a}. 
The cumulant
generating function is given by $S(\chi
,t)=\ln\big\{\textrm{Tr}[\exp(\mathcal{L}(\chi)t)\bar\rho]\big\}$ with
steady-state density matrix $\bar\rho$ given by $\mathcal{L}(0)\bar\rho =0$. The
cumulants are obtained via $C_k(t)=(-i)^k\partial_\chi^k S(\chi
,t)\big|_{(\chi =0)}$. The steady-state current is $\langle
I\rangle =e\dot C_1$ and the zero-frequency Fano factor
(DC noise) is defined as $F\equiv \dot C_2/\dot C_1$. The Fano factor
indicates sub(super)-Poissonian electron transfer if it is
smaller(larger) than unity. 

\begin{figure}[t]
\includegraphics[width=.48\textwidth,clip=true]{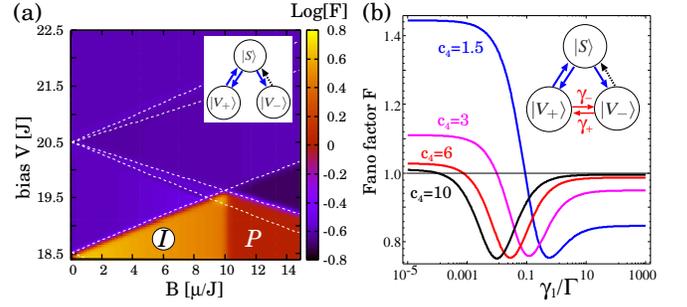}
\caption{(color online). (a) Fano factor {\em vs.} symmetric bias voltage $V$ and magnetic
field $B$, parameters as in Fig.~\ref{fig1}c; region $P$: Poissonian
charge transfer; Inset: configuration
space with single-electron tunneling transitions for region $I$ 
in Fig.~\ref{fig1}c, Transition $|V_-\rangle\rightarrow| S\rangle$ is exponentially suppressed. (b) Fano
factor {\em vs.} spin decay rate $\gamma_1$ for symmetric coupling $\Gamma
=\frac{1}{2}\Gamma_{L/R}$ and various coefficients $c_4\propto
B$ in region $I$ (see Fig.~\ref{fig2}a); Inset: Configuration
space with indicated spin flip rates. 
}
\label{fig2}
\end{figure}

The spectrum of $H_\textrm{QD}$ (Tab.~\ref{tab1}) itself is not directly
accessible by transport spectroscopy. Rather, by applying a bias voltage $V$
to the SET, the excitation spectrum is probed. The six transitions with nonvanishing Clebsch-Gordan
coefficients are indicated in
Fig.~\ref{fig1}b. Fig.~\ref{fig1}c shows that the current {\em vs.} magnetic field $B$ and bias
voltage $V$ is a direct map of those transitions. The six steps correspond to excitations in the
eigenspectrum (assuming the temperature is low, otherwise the
transitions would be smeared out). Between steps the current is nearly
constant for fixed, but changes with variable magnetic field due to the
dependence of the Clebsch-Gordan coefficients. At
zero-magnetic field, two current steps occur corresponding to tunneling
transitions from the unoccupied states to the singlet- and degenerate
triplet states. The bias
difference of the steps is $2J$. From the positions of the current
steps one easily obtains the g-factors of both spins by fitting the
excitation energies calculated by the differences of eigenenergies in Tab.~\ref{tab1}. We point out that the 
specific dependence of the excitation spectrum on the magnetic field
can appear very different when one of the g-factors is negative (as
widely occurs in confined electron systems) e.g. for $g_1<0$ negative differential
conductance appears for certain transitions.

Remarkably, certain transitions are completely missing (dotted lines
in Fig.~\ref{fig1}c)
even though the corresponding Clebsch-Gordan coefficient is not
zero. This is to be contrasted with spin blockade for which the
relevant coefficients disappear \cite{WEI95}. This blockade here is caused by the topology of the
configuration space. Let us focus on the lowest lying transition in
Fig.~\ref{fig1}c where
the configuration space consists of three states: $|V_+\rangle$, $|V_-\rangle$, and $| S\rangle$ 
in the inset of Fig.~\ref{fig2}a. Since the
transition energy for $|V_-\rangle\rightarrow|
S\rangle$ is too large to be supplied by the lead electrons in that
bias range, the rate for that
process is exponentially suppressed. Consequently, the system gets
trapped in the state $|V_-\rangle$ once
it reaches there, the spin $S_1$ is completely polarized and the
current is blocked. Such population trapping effect can
be expected in any system with a configuration space with one or more
singly-connected states. We emphasize that the current blockade persists for
an arbitrary choice of $J$, $g_{1,2}$, $S_1$, and even for
ferromagnetic leads 
$\Gamma_{L,R}^\uparrow\not=\Gamma_{L,R}^\downarrow$. However, 
the blockade will be removed with increasing temperature $T$ since the transition 
$|V_-\rangle\rightarrow|S\rangle$ becomes more likely (compare inset of Fig.~\ref{fig2}a). 
In regions $I$
and $II$ the dynamics can be described by an effective three-state
model assuming low temperatures such that $f_R(\epsilon_{s1})=f_R(\epsilon_{s2})=0$ and excitations
involving triplet states are negligible:

\begin{eqnarray}
\dot \rho_{|\downarrow
0\rangle}&=&\{-\Gamma_Lf_L(\epsilon_{s1})\rho_{|\downarrow
0\rangle}+[\Gamma_Lf_L^-(\epsilon_{s1})+\Gamma_R]\rho_{|S\rangle}\}c_2^{-2},\nonumber\\
\dot \rho_{|\uparrow 0\rangle}&=&\{-\Gamma_Lf_L(\epsilon_{s2})\rho_{|\uparrow
0\rangle}+[\Gamma_Lf_L^-(\epsilon_{s2})+\Gamma_R]\rho_{|S\rangle}\}\frac{c_4^{2}}{c_2^2},\nonumber\\
\dot \rho_{|S\rangle}&=&\big\{\Gamma_Lf_L(\epsilon_{s1})\rho_{|\downarrow
0\rangle}+c_4^2\Gamma_Lf_L(\epsilon_{s2})\rho_{|\uparrow
0\rangle}\nonumber\\
&-&\big[\Gamma_L[f_L^-(\epsilon_{s1})+c_4^2f_L^-(\epsilon_{s2})]+\Gamma_Rc_2^{2}\big]\rho_{|S\rangle}\big\}c_2^{-2}
\end{eqnarray}
with $\epsilon_{s1/s2}\equiv\varepsilon_2\pm\varepsilon_z^{(1)}/2-1/4-c/2$
and $f_L^-(\cdot )\equiv 1-f_L(\cdot )$.

The Fano factor in region $I$ is
$F_I=1+2c_4^2\Gamma_R /(\Gamma_L +\Gamma_R)$
and clearly indicates super-Poissonian electron transfer since $c_4>1$
for $B>0$ (lower left corner of Fig.~\ref{fig2}a).  A very similar expression has been derived in Ref.~\cite{BOD08} with the interaction parameter 
$\varPhi$ replacing $c_4$.
The
underlying mechanism is related to thermally-activated bunching of tunneling
events \cite{BEL05}: as long as the system is in state $|V_-\rangle$
no tunneling occurs. However, with an exponentially small probability,
thermally-excited lead electrons can enter the state $| S\rangle$
and a small bunch of tunneling transitions $|V_+\rangle\leftrightarrow |
S\rangle$ may take place. Outside of region $I$ the noise is
Poissonian ($P$)
or sub-Poissonian as can be seen in Fig.~\ref{fig2}a for symmetric
lead couplings $\Gamma_L=\Gamma_R$. For asymmetric couplings, it is
also possible to observe $F>1$ in other regions, e.g., in region $II$
when the condition $\Gamma_L/\Gamma_R > (c_4-1)^2/\, 2c_4$ is fulfilled. 

We can exploit the current blockade mechanism in order to measure the spin
relaxation rate $\gamma_1=1/T_1$ of $S_1$. In the inset of
Fig.~\ref{fig2}b we show that the formerly singly-connected state
$|V_-\rangle$ is additionally linked with state
$|V_+\rangle$ via spin flip transitions with rates 
$\gamma_+=\gamma_1f\big(\varepsilon_z^{(1)}\big)$,
$\gamma_-=\gamma_1\big[1-f\big(\varepsilon_z^{(1)}\big)\big]$. Such a mono-exponential spin decay can occur, e.g., due to 
spin-orbit coupling \cite{KRO04}.
The
steady-state current is then directly related to the spin-relaxation rate
$1/T_1$ and it can be simply measured by the ratio of plateau  currents in
regions $I$ and $II$ at the same magnetic field	(assuming
$\gamma_1\ll\Gamma$ so that $\langle I\rangle_{II}$ does not depend
on $\gamma_1$):

\begin{eqnarray}
\gamma_1^{-1} &=& A_1 \bigg[ A_2\frac{\langle I\rangle_{II}}{\langle I\rangle_{I}}-A_3\bigg]
\end{eqnarray}
with $A_1\equiv c_2^2/(\Gamma\Gamma_Lc_4^2)$, $A_2\equiv [1-f\big(\varepsilon_z^{(1)}\big)](\Gamma_L+2\Gamma_R)$,
$A_3\equiv (c_4^2\Gamma
+\Gamma_L[1-f\big(\varepsilon_z^{(1)}\big)]+\Gamma_R)$,  and $\Gamma\equiv\Gamma_L+\Gamma_R$.

The Fano factor {\em vs.} $\gamma_1$ in Fig.~\ref{fig2}b shows that by
increasing the spin relaxation the charge transfer turns
from super- to sub-Poissonian. For large magnetic fields the Fano
factor approaches unity: for small $\gamma_1$ from the
super-Poissonian side and for large $\gamma_1$ from the sub-Poissonian
side. Hence, the zero-frequency noise provides a sensitive
quantitive indicator of the spin decay.

In region $I$ the exponential suppression of the current can also be
lifted by algebraic cotunneling contributions \cite{AVE90}.
To ensure that the effect of spin relaxation on the SET transport is
not obscured by higher-order tunneling processes, the rates $\Gamma_{L/R}$
have to be small with respect to the temperature.

The discussed effects are robust against background charge
fluctuations. We have checked that by considering fluctuating energy
levels \mbox{$\varepsilon_i(t)=\varepsilon_i +1/\sqrt{\tau_i}\,\xi_i(t)$} ($i=1,2$) with
\mbox{$\langle\xi_i(t)\rangle =0$ and $\langle\xi_i(t)\xi_j(t')\rangle =\delta_{ij}\delta (t-t')$}.
Second-order perturbation theory in the fluctuation strength leads to
the Lindblad form \cite{BRE02}
\mbox{$\mathcal{L}_{\textrm{cf}}\rho =\sum_i\tau_i^{-1}[n_i\rho n_i
-\frac{1}{2}\{n_i^2,\rho\}]$}, which leaves the relevant subspace invariant.

In summary, we have studied a Coulomb-blockaded SET with exchange-coupled 1/2-spin in a
constant magnetic field. The
excitation spectrum (current {\em vs.} magnetic field and bias voltage)
shows rich structure enabling the measurement of the exchange-coupling
strength $J$ and the Land\'e g-factors. Due to classical population trapping
the lowest-lying transition is blocked independent of $J$, the
g-factors and the spin size - the current is exponentially
suppressed and the electron transfer is super-Poissonian. As we
demonstrate, this robust phenomenon allows the direct measurement of
the decay rate $1/T_1$ of the probed spin  by DC current and noise
measurement. 
Our method can be readily used to implement larger $S_1$ encountered in
nanomagnets or to describe the nuclear spin environment of a QD.

Helpful discussions with W. Belzig are gratefully acknowledged.
This work was supported financially by the Deutsche Forschungsgemeinschaft in project \mbox{BR 1528/5-1}.



\begin{thebibliography}{13}
\expandafter\ifx\csname natexlab\endcsname\relax\def\natexlab#1{#1}\fi
\expandafter\ifx\csname bibnamefont\endcsname\relax
  \def\bibnamefont#1{#1}\fi
\expandafter\ifx\csname bibfnamefont\endcsname\relax
  \def\bibfnamefont#1{#1}\fi
\expandafter\ifx\csname citenamefont\endcsname\relax
  \def\citenamefont#1{#1}\fi
\expandafter\ifx\csname url\endcsname\relax
  \def\url#1{\texttt{#1}}\fi
\expandafter\ifx\csname urlprefix\endcsname\relax\def\urlprefix{URL }\fi
\providecommand{\bibinfo}[2]{#2}
\providecommand{\eprint}[2][]{\url{#2}}

\bibitem[{\citenamefont{Loss and Vincenzo}(1998)}]{LOS98}
\bibinfo{author}{\bibfnamefont{D.}~\bibnamefont{Loss}} \bibnamefont{and}
  \bibinfo{author}{\bibfnamefont{D.~P.} \bibnamefont{Vincenzo}},
  \bibinfo{journal}{Phys.~Rev.~A} \textbf{\bibinfo{volume}{57}},
  \bibinfo{pages}{120} (\bibinfo{year}{1998});
\bibinfo{author}{\bibfnamefont{G.}~\bibnamefont{Burkhard}},  
\bibinfo{author}{\bibfnamefont{D.}~\bibnamefont{Loss}},  
\bibnamefont{and}
  \bibinfo{author}{\bibfnamefont{D.~P.} \bibnamefont{Vincenzo}},
  \bibinfo{journal}{Phys.~Rev.~B} \textbf{\bibinfo{volume}{59}},
  \bibinfo{pages}{2070} (\bibinfo{year}{1999}).

\bibitem[{\citenamefont{Elzerman et~al.}(2004)\citenamefont{Elzerman, Hanson,
  van Beveren, Witkamp, Vandersypen, and Kouwenhoven}}]{ELZ04}
\bibinfo{author}{\bibfnamefont{J.~M.} \bibnamefont{Elzerman}},
  \bibinfo{author}{\bibfnamefont{R.}~\bibnamefont{Hanson}},
  \bibinfo{author}{\bibfnamefont{L.~H.~W.} \bibnamefont{van Beveren}},
  \bibinfo{author}{\bibfnamefont{B.}~\bibnamefont{Witkamp}},
  \bibinfo{author}{\bibfnamefont{L.~M.~K.} \bibnamefont{Vandersypen}},
  \bibnamefont{and} \bibinfo{author}{\bibfnamefont{L.~P.}
  \bibnamefont{Kouwenhoven}},
  \bibinfo{journal}{Nature}
  \textbf{\bibinfo{volume}{430}}, \bibinfo{pages}{431} (\bibinfo{year}{2004}).

\bibitem[{\citenamefont{Wabnig et~al.}(2009)\citenamefont{Wabnig, Lovett,
  Jefferson, and Briggs}}]{WAB09}
\bibinfo{author}{\bibfnamefont{J.}~\bibnamefont{Wabnig}},
  \bibinfo{author}{\bibfnamefont{B.~W.} \bibnamefont{Lovett}},
  \bibinfo{author}{\bibfnamefont{J.~H.} \bibnamefont{Jefferson}},
  \bibnamefont{and} \bibinfo{author}{\bibfnamefont{G.~A.~D.}
  \bibnamefont{Briggs}},
  \bibinfo{journal}{Phys.~Rev.~Lett.}
  \textbf{\bibinfo{volume}{102}}, \bibinfo{pages}{016802}
  (\bibinfo{year}{2009}).

\bibitem[{\citenamefont{Yoo et~al.}(1997)\citenamefont{Yoo, Fulton, Hess,
  Willett, Dunkleberger, Chichester, Pfeiffer, and West}}]{YOO97}
\bibinfo{author}{\bibfnamefont{M.~J.} \bibnamefont{Yoo}},
  \bibinfo{author}{\bibfnamefont{T.~A.} \bibnamefont{Fulton}},
  \bibinfo{author}{\bibfnamefont{H.~F.} \bibnamefont{Hess}},
  \bibinfo{author}{\bibfnamefont{R.~L.} \bibnamefont{Willett}},
  \bibinfo{author}{\bibfnamefont{L.~N.} \bibnamefont{Dunkleberger}},
  \bibinfo{author}{\bibfnamefont{R.~J.} \bibnamefont{Chichester}},
  \bibinfo{author}{\bibfnamefont{L.~N.} \bibnamefont{Pfeiffer}},
  \bibnamefont{and} \bibinfo{author}{\bibfnamefont{K.~W.} \bibnamefont{West}},
  \bibinfo{journal}{Science} \textbf{\bibinfo{volume}{276}},
  \bibinfo{pages}{579} (\bibinfo{year}{1997}).

\bibitem[{\citenamefont{Rugar et~al.}(2004)\citenamefont{Rugar, Budakian,
  Mamin, and Chui}}]{RUG04}
\bibinfo{author}{\bibfnamefont{D.}~\bibnamefont{Rugar}},
  \bibinfo{author}{\bibfnamefont{R.}~\bibnamefont{Budakian}},
  \bibinfo{author}{\bibfnamefont{H.~J.} \bibnamefont{Mamin}}, \bibnamefont{and}
  \bibinfo{author}{\bibfnamefont{B.~W.} \bibnamefont{Chui}},
  \bibinfo{journal}{Nature} \textbf{\bibinfo{volume}{430}},
  \bibinfo{pages}{329} (\bibinfo{year}{2004}).

\bibitem[{\citenamefont{Fernandez-Rossier and Aguado}(2007)}]{ROS07}
\bibinfo{author}{\bibfnamefont{J.}~\bibnamefont{Fernandez-Rossier}}
  \bibnamefont{and} \bibinfo{author}{\bibfnamefont{R.}~\bibnamefont{Aguado}},
  \bibinfo{journal}{Phys.~Rev.~Lett.} \textbf{\bibinfo{volume}{98}},
  \bibinfo{pages}{106805} (\bibinfo{year}{2007}).


\bibitem[{\citenamefont{Ono and Tarucha}(2004)}]{ONO04}
\bibinfo{author}{\bibfnamefont{K.}~\bibnamefont{Ono}} \bibnamefont{and}
  \bibinfo{author}{\bibfnamefont{S.}~\bibnamefont{Tarucha}},
  \bibinfo{journal}{Phys.~Rev.~Lett.} \textbf{\bibinfo{volume}{92}},
  \bibinfo{pages}{256803} (\bibinfo{year}{2004}).

\bibitem[{\citenamefont{I{\~n}arrea et~al.}(2007)\citenamefont{I{\~n}arrea,
  Platero, and MacDonald}}]{INA07a}
\bibinfo{author}{\bibfnamefont{J.}~\bibnamefont{I{\~n}arrea}},
  \bibinfo{author}{\bibfnamefont{G.}~\bibnamefont{Platero}}, \bibnamefont{and}
  \bibinfo{author}{\bibfnamefont{A.-H.} \bibnamefont{MacDonald}},
  \bibinfo{journal}{Phys.~Rev.~B} \textbf{\bibinfo{volume}{76}},
  \bibinfo{pages}{085329} (\bibinfo{year}{2007}).

\bibitem[{\citenamefont{Golovach and Loss}(2004)}]{GOL04}
\bibinfo{author}{\bibfnamefont{V. N.}~\bibnamefont{Golovach}} \bibnamefont{and}
  \bibinfo{author}{\bibfnamefont{D.}~\bibnamefont{Loss}},
  \bibinfo{journal}{Phys.~Rev.~B} \textbf{\bibinfo{volume}{69}},
  \bibinfo{pages}{245327} (\bibinfo{year}{2004}).

\bibitem[{\citenamefont{Wabnig and Lovett}(2009)}]{WAB09a}
\bibinfo{author}{\bibfnamefont{J.}~\bibnamefont{Wabnig}} \bibnamefont{and}
  \bibinfo{author}{\bibfnamefont{B. W.}~\bibnamefont{Lovett}},
  \bibinfo{journal}{New J. Phys.} \textbf{\bibinfo{volume}{11}},
  \bibinfo{pages}{043031} (\bibinfo{year}{2009}).

\bibitem[{\citenamefont{Breuer and Petruccione}(2002)}]{BRE02}
\bibinfo{author}{\bibfnamefont{H.-P.} \bibnamefont{Breuer}} \bibnamefont{and}
  \bibinfo{author}{\bibfnamefont{F.}~\bibnamefont{Petruccione}},
  \emph{\bibinfo{title}{The theory of open quantum systems}}
  (\bibinfo{publisher}{Oxford University Press}, \bibinfo{address}{Great
  Clarendon Street}, \bibinfo{year}{2002}).

\bibitem[{\citenamefont{Schaller and Brandes}(2008)}]{SCH08}
\bibinfo{author}{\bibfnamefont{G.}~\bibnamefont{Schaller}} \bibnamefont{and}
  \bibinfo{author}{\bibfnamefont{T.}~\bibnamefont{Brandes}},
  \bibinfo{journal}{Phys.~Rev.~A} \textbf{\bibinfo{volume}{78}},
  \bibinfo{pages}{022106} (\bibinfo{year}{2008}).

\bibitem[{\citenamefont{Bagrets and Nazarov}(2003)}]{BAG03a}
\bibinfo{author}{\bibfnamefont{D.~A.} \bibnamefont{Bagrets}} \bibnamefont{and}
  \bibinfo{author}{\bibfnamefont{Y.~V.} \bibnamefont{Nazarov}},
  \bibinfo{journal}{Phys.~Rev.~B} \textbf{\bibinfo{volume}{67}},
  \bibinfo{pages}{085316} (\bibinfo{year}{2003});
\bibinfo{author}{\bibfnamefont{G.} \bibnamefont{Schaller}},
  \bibinfo{author}{\bibfnamefont{G.} \bibnamefont{Kiesslich}},
\bibnamefont{and}
  \bibinfo{author}{\bibfnamefont{T.} \bibnamefont{Brandes}},
  (\bibinfo{year}{unpublished, arXiv:0908.3620});

\bibitem[{\citenamefont{Weinmann et~al.}(1995)\citenamefont{Weinmann,
  H{\"a}usler, and Kramer}}]{WEI95}
\bibinfo{author}{\bibfnamefont{D.}~\bibnamefont{Weinmann}},
  \bibinfo{author}{\bibfnamefont{W.}~\bibnamefont{H{\"a}usler}},
  \bibnamefont{and} \bibinfo{author}{\bibfnamefont{B.}~\bibnamefont{Kramer}},
  \bibinfo{journal}{Phys.~Rev.~Lett.} \textbf{\bibinfo{volume}{74}},
  \bibinfo{pages}{984} (\bibinfo{year}{1995}).

\bibitem[{\citenamefont{Bodoky et~al.}(2008)\citenamefont{Bodoky,
  Belzig, and Bruder}}]{BOD08}
\bibinfo{author}{\bibfnamefont{F.}~\bibnamefont{Bodoky}},
  \bibinfo{author}{\bibfnamefont{W.}~\bibnamefont{Belzig}},
  \bibnamefont{and} \bibinfo{author}{\bibfnamefont{C.}~\bibnamefont{Bruder}},
  \bibinfo{journal}{Phys.~Rev.~B} \textbf{\bibinfo{volume}{77}},
  \bibinfo{pages}{035302} (\bibinfo{year}{2008}).

\bibitem[{\citenamefont{Belzig}(2005)}]{BEL05}
\bibinfo{author}{\bibfnamefont{W.}~\bibnamefont{Belzig}},
  \bibinfo{journal}{Phys.~Rev.~B} \textbf{\bibinfo{volume}{71}},
  \bibinfo{pages}{161301(R)} (\bibinfo{year}{2005}).

\bibitem[{\citenamefont{Kroutvar et~al.}(2004)}]{KRO04}
\bibinfo{author}{\bibfnamefont{M.}~\bibnamefont{Kroutvar}},
\bibinfo{author}{\bibfnamefont{Y.}~\bibnamefont{Ducommun}}, 
\bibinfo{author}{\bibfnamefont{D.}~\bibnamefont{Heiss}}, 
\bibinfo{author}{\bibfnamefont{M.}~\bibnamefont{Bichler}}, 
\bibinfo{author}{\bibfnamefont{D.}~\bibnamefont{Schuh}}, 
\bibinfo{author}{\bibfnamefont{G.}~\bibnamefont{Abstreiter}}, 
\bibnamefont{and}
\bibinfo{author}{\bibfnamefont{J. F.}~\bibnamefont{Finley}},
  \bibinfo{journal}{Nature} \textbf{\bibinfo{volume}{432}},
  \bibinfo{pages}{81} (\bibinfo{year}{2004}).

\bibitem[{\citenamefont{Averin, Nazarov}(1990)}]{AVE90}
\bibinfo{author}{\bibfnamefont{D. V.}~\bibnamefont{Averin}} \bibnamefont{and}
\bibinfo{author}{\bibfnamefont{Yu. V.}~\bibnamefont{Nazarov}},
  \bibinfo{journal}{Phys.~Rev.~Lett.} \textbf{\bibinfo{volume}{65}},
  \bibinfo{pages}{2446} (\bibinfo{year}{1990}).

\end{thebibliography}
\end{document}